\documentclass[aps,pra,longbibliography,showkeys,showpacs,groupedaddress,twocolumn,10pt]{revtex4-1}
\usepackage{amssymb}
\usepackage{graphicx}
\usepackage{dcolumn}
\usepackage{bm}
\usepackage{amsmath}
\usepackage{subfigure}
\usepackage{float}
\usepackage{color}

\begin{document}

\title{Atom-based quantum receiver for amplitude- and frequency-modulated baseband signals in high-frequency radio communication}

\author{Yuechun Jiao$^{1,3}$}
\author{Xiaoxuan Han$^{1,3}$}
\author{Jiabei Fan$^{1,3}$}
\author{Georg Raithel$^{1,2}$}
\author{Jianming Zhao$^{1,3}$}
\thanks{Corresponding author: zhaojm@sxu.edu.cn}
\author{Suotang Jia$^{1,3}$}

\affiliation{$^{1}$State Key Laboratory of Quantum Optics and Quantum Optics Devices, Institute of Laser Spectroscopy, Shanxi University, Taiyuan 030006, China}
\affiliation{$^{2}$ Department of Physics, University of Michigan, Ann Arbor, Michigan 48109-1120, USA}
\affiliation{$^{3}$Collaborative Innovation Center of Extreme Optics, Shanxi University, Taiyuan 030006, China}

\begin{abstract}
An optical probe of cesium Rydberg atoms generated in a thermal vapor cell is used to retrieve a baseband signal modulated onto a 16.98-GHz carrier wave in real-time, demonstrating an atom-based quantum receiver suitable for microwave communication. The 60$S_{1/2}$ Rydberg level of cesium atoms in the cell is tracked via electromagnetically induced transparency (EIT), an established laser-spectroscopic method. The microwave carrier is resonant with the 60$S_{1/2}$ $\rightarrow$ 60$P_{1/2}$ Rydberg transition, resulting in an Autler-Townes (AT) splitting of the EIT signal. Amplitude modulation of the carrier wave results in a corresponding modulation in the optically retrieved AT splitting. Frequency modulation causes a change in relative height of the two AT peaks, which can be optically detected and processed to retrieve the modulation signal. The optical retrieval of the baseband signal does not require electronic demodulation. The method is suitable for carrier frequencies within a range from $\sim 1$~GHz to hundreds of GHz. The baseband bandwidth, which is $\sim$~20~Hz in the present demonstration, can be increased by faster spectroscopic sampling.
\end{abstract}
\keywords{Rydberg EIT, AT splitting, atom-based quantum sensor or receiver, microwave communication}
\pacs{32.80.Ee, 42.50.Gy, 84.40.-x, }
\maketitle

\section{Introduction}

Communication technologies play an ever increasing role in social life, science and defense, and are continuously being  developed further. Typically, communication is based on modulation/demondulation of a baseband signal onto an electromagnetic carrier wave via amplitude (AM) or frequency modulation (FM). For instance, in AM and FM radio broadcasting the carrier frequency ranges from hundreds of kHz to several hundred MHz.  The radio-wave receiving unit typically includes an antenna and a demodulation circuit.
A recent surge in the development of quantum technologies for atom-based field detection will naturally lead into novel atom-based transmitter and receiver components that are specialized for communications tasks.
Recently, atom-based quantum sensors for electro-magnetic fields have already been explored for classical communication~\cite{Gerginov}.

\begin{figure}[ht]
\centering
\includegraphics[width=0.45\textwidth]{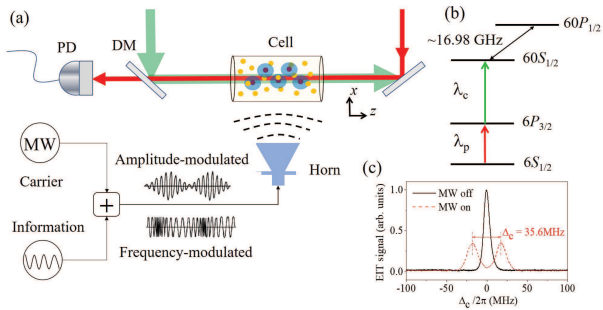}
\caption{(a) Schematic of the experimental setup. The coupling and probe beams, with respective wavelengths
$\lambda_{c}$ and $\lambda_{p}$, counter-propagate through and overlap within a cesium vapor cell. The probe beam is passed through the cell and a dichroic mirror (DM), and is detected with a photodiode (PD), yielding the probe transmission.
The microwave field (MW) has a carrier frequency of 16.98~GHz and is amplitude- or frequency-modulated with a low-frequency
baseband signal; the modulation contains the transmitted information. The modulated MW signal is fed into a horn antenna and radiated into the atomic cell, which receives the signal. The laser and microwave electric fields have the same polarization, transverse to the drawing plane.  (b) Four-level Rydberg-EIT scheme. The microwave field is (near-)resonant with the 60$S_{1/2}$ $\rightarrow$ 60$P_{1/2}$ Rydberg transition. The resultant AT splitting is measured and used as a real-time, optical readout for the modulation signal. (c) Samples of measured EIT spectra with MW off (black solid line) and on (red dashed line) vs coupler-laser detuning, $\Delta_c$. In the displayed on-resonant case, the AT splitting observed with the MW on, $f_{AT}$, approximately equals the Rabi frequency of the MW transition, $\Omega$.}
\end{figure}

Rydberg atoms, highly excited atoms with principal quantum numbers $n \gg 1$, can be used to create sensitive electric field sensors, due to their large polarizabilities ($\propto n^{7}$) and microwave-transition dipole moments ($\propto n^2$)~\cite{Gallagher}. Due to the invariant nature of atomic properties, Rydberg-atom-based field metrology has clear advantages over traditional electrometry, because of SI traceability and potentially calibration-free operation~\cite{C. Holloway2014, C. Holloway2017}. Rydberg electromagnetically induced transparency (Rydberg-EIT)~\cite{Mohapatra} in vapor cell has been employed to measure the properties of electric fields over a wide frequency range from 100~MHz~\cite{Jiao} to over 1~THz~\cite{H. Fan}, including measurements of microwave (MW) fields~\cite{Sedlacek2012} and their polarizations~\cite{Sedlacek2013}, and millimeter waves~\cite{Gordon}.
Small Rydberg-atom field sensors that employ $\mu$m-length
vapor cells and hollow-core fibers~\cite{Veit2016} offer significant potential for miniaturization.
Recently, Rydberg atoms have also been explored as sensitive, high bandwidth, atomic communications receivers for digital communication~\cite{Meyer}. Rydberg atoms have many electronic states with a large number of electric-dipole transitions between them~\cite{Gallagher}, leading to a strong electromagnetic response of these atoms at a dense set of frequencies within the MHz- to THz-range. Rydberg-atom-based field detectors can have a higher sensitivity than detectors with traditional dipole antennas~\cite{Kumar}, making them suitable for long-distance communication with potential for high-speed parallel operation. In addition, Rydberg atomic receivers can be used for subwavelength imaging of microwave electric-field distributions~\cite{Fan, Holloway}.

In this work, we propose a quantum sensor for microwave communication based on Rydberg atoms in a room-temperature cell, which can enable real-time and direct readout of information encoded onto a high-frequency carrier. In our experimental demonstration, we encode a low-frequency baseband signal (bandwidth in the 10-Hz range) on a 16.98~GHz microwave carrier via amplitude or frequency modulation. The high-frequency carrier resonantly interacts with the cesium 60$S_{1/2}$ $\rightarrow$ 60$P_{1/2}$ Rydberg transition, and the encoded information is recovered by employing the Rydberg electromagnetically-induced-transparency and the Autler-Townes effect (Rydberg-EIT-AT). This constitutes a quantum sensor and receiver element for microwave communication with a carrier frequency in the tens of GHz range. Extension to carrier frequencies in the hundreds of MHz to 1~THz range is straightforward.

\section{Experimental Setup}

A schematic of the experimental setup and the relevant four-level system are shown in Figs.~1~(a) and (b). The experiments are performed in a room-temperature cesium vapor cell. The coupling and probe beams are overlapped and counter-propagated along the centerline of the cell. A 852-nm ($\lambda_{p}$) probe laser, resonantly interacting with the lower transition, $|6 S_{1/2}, F=4\rangle $ $\to$ $|6P_{3/2}, F'=5\rangle $ with Rabi frequency $\Omega_p$ = 2$\pi \times $12.3~MHz, and beam waist $w_0 = 105~\mu$m, and a 510-nm ($\lambda_{c}$) coupling beam, scanning through the Rydberg transition $|6P_{3/2}, F'=5 \rangle $ $\to$ $|60S_{1/2}\rangle$ with $\Omega_c$ = 2$\pi \times $2.45~MHz and $w_0= 150~\mu$m, form the Rydberg-EIT system that is employed to measure the modulated microwave field.
The power of the probe beam passing through the cell is detected with a photodiode and recorded with an oscilloscope. This yields an all-optical readout for the field strength and the frequency of the microwave field as a function of time.

The MW field, generated with a signal generator (Keysight N5183) and emitted from a horn antenna, is applied transversely to the laser beams propagating through the vapor cell, where it interacts with cesium receiver Rydberg atoms. A baseband signal function is amplitude- or frequency-modulated onto the carrier microwave.
The MW carrier has a frequency of 16.98~GHz, which drives the 60$S_{1/2}$-60$P_{1/2}$ Rydberg transition with an on-resonance Rabi frequency $\Omega$. The resultant EIT-AT splitting spectra, a sample of which is shown in Fig.~1(c), are used as a real-time optical readout of the modulation signal that is encoded in the MW.

\begin{figure}[thb]
\centering
\includegraphics[width=0.4\textwidth]{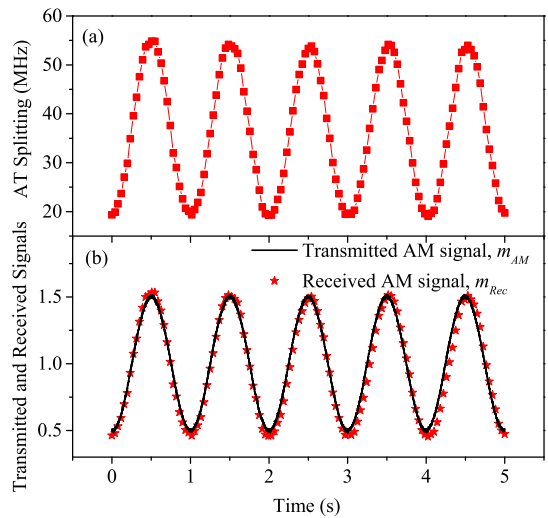}
\vspace{-1ex}
\caption{Amplitude modulation results for sinusoidal modulation. (a) Measurement of the AT splitting, $f_{AT}(t)$.
(b) Comparison between the amplitude modulation function applied to the microwave generator, $m_{AM}(t)$ (transmitted
baseband signal; black line),
and the corresponding received signal, $m_{Rec}=E_{cm}(t)/E_c$ (stars), which is derived from the spectroscopically acquired
AT splitting shown in (a). For the case shown, $\omega_s = 2 \pi \times 1$~Hz,  $E_c = 2.25$~V/m and $m_{AM}(t) = 1 - 0.50 \cos(\omega_s t)$.}
\end{figure}

\section{Amplitude modulation}

In our initial demonstration, the signal to be transmitted is an amplitude-modulated signal function
$m_{AM}(t) = 1 - k_{AM} \cos(\omega_s t)$ with a signal frequency $\omega_s$ and modulation index $k_{AM}$.
Denoting the carrier frequency $\omega_c$ and the fixed (unmodulated) field amplitude $E_c$, the modulated MW field impinging onto the atoms in the vapor cell is
\begin{equation}
E_{AM}(t) =  E_c m_{AM}(t) \cos(\omega_c t).
\end{equation}

The modulated field amplitude, $E_{cm}(t) = E_c m_{AM}(t)$ $= E_c (1 - k_{AM}\cos(\omega_s t))$, is proportional to the signal function $m_{AM}(t)$ to be transmitted. On the receiver end, the objective therefore is to measure $E_{cm}(t)$ by rapid acquisition and analysis of Rydberg-EIT-AT spectra. A sample of such a spectrum is shown in Fig.~1(c). Denoting the measured slowly-varying AT splitting frequency by $f_{AT}$, we retrieve $E_{cm}(t)$ using the relation
\begin{equation}
E_{cm}(t)  =  2\pi\frac{\hbar}{\wp}f_{AT}(t)
\label{eq:AT}
\end{equation}
where $\hbar$ is Planck's constant and $\wp$ is the atomic dipole moment of the MW transition (here, $\wp=1.046 \times 10^{-26}$~Cm).
For Eq.~\ref{eq:AT} to approximately hold, the carrier frequency must be resonant with the desired Rydberg transition (here 60$S_{1/2}$-60$P_{1/2}$), and other conditions apply~\cite{C. Holloway2017}. Also, to recover the signal function $m_{AM}(t)$ in real time, the spectrum acquisition time must be much shorter than $2 \pi/\omega_s$.

In Fig.~2(a) we show the measured AT splitting, $f_{AT}(t)$, which yields the MW field amplitude $E_{cm}(t)$ according to
Eq.~\ref{eq:AT}. For the case in Fig.~2(a), the modulation-free electric field of the carrier is determined to be 2.25~V/m.
When the modulation is applied, the sinusoidal oscillation of the detected MW field normalized by the carrier field, $m_{Rec}(t) = E_{cm}(t)/E_c$, directly reproduces the applied modulation function $m_{AM}(t)$.
To show that the retrieved signal $m_{Rec}(t)$ equals the applied modulation signal $m_{AM}(t)$, in Fig.~2(b)
we compare both functions. Overall we observe good agreement, as expected.  We determine the deviation between the transmitted and received signals, $\vert m_{AM}(t) - m_{Rec}(t) \vert$, to be $\lesssim 5\%$ of the average
transmitted signal and a fidelity of the recovered signal of $\gtrsim$ 95$\%$. 

For the case of sinusoidal modulations, data as shown in Fig.~2 also allow rapid retrieval of the modulation frequency, $\omega_s$, and the modulation index. For instance, in Fig.~2 the minima and maxima of the retrieved $E_{cm}(t)$ are $E_{cm,min}$ =1.13 V/m and $E_{cm,max}$ = 3.38 V/m, yielding a reading for the modulation index of the transmitted and received signals, $m_{AM}(t)$ and $m_{Rec}(t)$, of
\begin{eqnarray}
k_{AM} \approx \frac{E_{cm,max} - E_{cm,min}}{E_{cm,max} + E_{cm,min}} \quad,
\end{eqnarray}
which is $k_{AM} = 0.50$ in Fig.~2.

\begin{figure}[thb]
\centering
\includegraphics[width=0.4\textwidth]{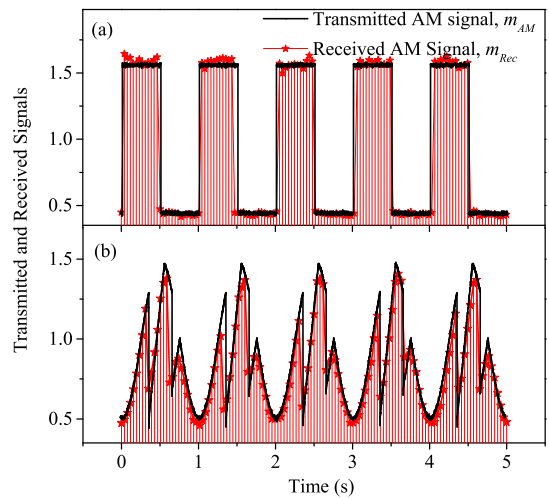}
\vspace{-1ex}
\caption{Amplitude modulation results for square (a) and complex (b) baseband modulation functions. Black solid lines show the transmitted functions,  $m_{AM}(t)$, and stars with vertical drop lines show the received signals, $m_{Rec}(t) = E_{cm}(t)/E_c$,  derived from spectroscopically acquired AT splittings.}
\end{figure}

To test the time resolution and the transmission bandwidth of the method, we have also investigated square and structure-rich amplitude modulation functions $m_{AM}(t)$. These have a repetition frequency of 1~Hz, with a much higher signal bandwidth (because of the sharp features in the functions). Fig.~3 shows that the transmitted and received modulation functions agree very well in both cases. From the lag and the deviations seen at the steps, we estimate a transmission bandwidth in the baseband of $\sim 20$~Hz, for the cases shown.

\section{Frequency modulation}

In radio broadcasting, FM modulation is often preferable to AM modulation because FM offers a higher bandwidth and is more resilient to interference and carrier-strength fluctuations. In this section we show that  Rydberg EIT-AT spectroscopy also can be used as a quantum receiver to retrieve information in FM-modulated signals.

In FM-mode, the microwave field has a fixed amplitude $E_c$ and a time-dependent phase $\Phi$ that carries the signal,
\begin{equation}
 E_{FM}(t) = E_{c} \cos(\omega_c t + \Phi(t))
 \end{equation}
with the phase modulation signal $\Phi(t)$. The transmitted FM signal is the time derivative of the phase, $\delta \omega_{FM}(t) =  \Phi'(t)$. Typically the baseband signal to be sent is proportional to $\delta \omega_{FM}(t)$. For instance, for a pure acoustic tone in the baseband, the function $\delta \omega_{FM}(t)$ varies sinusoidally at the frequency of the tone, and the amplitude of the function $\delta \omega_{FM}(t)$ is proportional to the square root of the sound intensity. For the present discussion, the FM-signal retrieval task amounts to extracting the transmitted function $\delta \omega_{FM}(t)$ from experimentally acquired Rydberg-EIT-AT spectra.

\begin{figure} [thb]
\vspace{-3ex}
\centering
\includegraphics[width=0.4\textwidth]{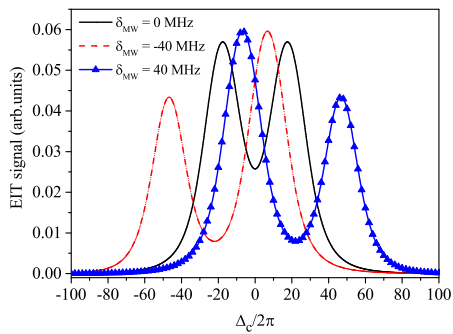}
\caption{Calculated Rydberg-EIT-AT spectra for MW fields interacting with the 60$S_{1/2}$ $\rightarrow$ 60$P_{1/2}$ Rydberg transition. The Rabi frequency of the MW transition is $\Omega=2 \pi \times 40~$MHz, and the frequency detunings are $\delta\omega_{FM}=0$ (black solid line), +40~MHz (blue dashed line) and -40~MHz (red dot-dashed line), respectively. For $\delta \omega_{FM} = 0$ the EIT-AT spectra exhibit two symmetric AT lines, while for non-zero detuning the AT lines become asymmetric in height. The (signed) difference in height yields an unambiguous reading for $\delta \omega_{FM}$. }
\end{figure}

Our method relies on the fact that the shape of the Rydberg-EIT-AT spectra changes as a function of the FM baseband signal function $\delta \omega_{FM}$. In Fig.~4, we present calculated spectra for several fixed values of the detuning $\delta \omega_{FM}$ = 2$\pi \times$ 0~MHz and $\pm$~40~MHz, respectively. The carrier frequency $\omega_c$ is chosen such that for $\delta \omega_{FM} = 0$ the microwave field is resonant with the Rydberg transition used for the detection. In the present demonstration we use a carrier frequency $\omega_c = 16.98$~GHz, which is resonant with the
60$S_{1/2}$-60$P_{1/2}$ transition. For $\delta \omega_{FM} = 0$, the EIT-AT spectrum exhibits two symmetric lines. As the detuning $\delta \omega_{FM}$ is tuned away from resonance,
the line separation increases and the line heights become asymmetric. It is found, in particular, that the detuning $\delta \omega_{FM} (t)$ is a single-valued function of the relative line-height difference, $F = (H_+ - H_-)/(H_+ + H_-)$, where $H_+$ and $H_-$ are the heights of the higher- and lower-frequency Rydberg-EIT-AT peaks relative to the reading away from the AT peaks. Assuming that the AT-peak line strengths are proportional to the squares of the $60S_{1/2}$-components in the corresponding AT eigenstates, we find by a straightforward analysis that the transmitted baseband signal $\delta \omega_{FM} (t)$ can be retrieved from the function $F$ via the relation
\begin{equation}
\delta \omega_{Rec} (t) = \frac{- F(t) \Omega}{\sqrt{1-F(t)^2}} \quad,
\label{eq:FM}
\end{equation}
where $\Omega$ is the Rabi frequency of the microwave transition between the Rydberg levels. Ideally, $\delta \omega_{Rec} (t) = \delta \omega_{FM} (t)$.
Equation~\ref{eq:FM} is applicable if the transmitted FM modulation signal $\delta \omega_{FM} (t)$ varies slowly enough to allow for real-time acquisition of the Rydberg-EIT-AT spectrum and extraction of the quantity $F(t)$. The Rydberg-EIT-AT spectra
are acquired in rapid succession, and the line heights $H_+(t)$ and $H_-(t)$ are extracted in real-time. The function $F = (H_+ - H_-)/(H_+ + H_-)$ and Eq.~\ref{eq:FM} are then used to calculate the received detuning $\delta \omega_{Rec}(t)$. This completes the FM signal retrieval task. It is noted that the only requirement for the spectroscopic readings $H_+$ and $H_-$  is that they must be proportional to the EIT-induced change of the transmission of the atomic vapor for the probe light. Also, the Rabi frequency $\Omega$ in Eq.~\ref{eq:FM} is given by the modulation-free splitting between the AT peaks ({\sl{i.e.}} the splitting seen at $\delta \omega_{FM} = 0$).

To demonstrate retrieval of an FM signal, in Fig.~5(a) we apply a sinusoidal FM modulation (transmitted FM signal) to the microwave interacting with the atoms,  $\delta \omega_{FM}(t) = - A \cos (\omega_s t)$ with an FM modulation frequency $\omega_s = 2 \pi \times 1$~Hz and modulation amplitude $A = 2 \pi \times$~40~MHz. The objective of the FM signal retrieval then is to recover the function $\delta \omega_{FM}(t)$, and its parameters  $\omega_s$ and $A$, from the spectroscopic response of the atoms to the microwave. In the retrieval process, Rydberg-EIT-AT spectra are acquired and analyzed at a rate higher than $1/\omega_s$. In Fig.~5(a) we compare the transmitted FM signal $\delta \omega_{FM}(t)$ and the received signal $\delta \omega_{Rec}(t)$ obtained from Eq.~\ref{eq:FM}. The figure shows that $\delta \omega_{Rec}(t) \approx \delta \omega_{FM}(t) $, demonstrating successful atom-based FM reception of a 1-Hz sinusoid. In Fig.~5(b) we repeat the procedure for an arbitrary modulation function that has a higher signal bandwidth in the baseband.  We again see good agreement between $\delta \omega_{FM}(t)$ and $\delta \omega_{Rec}(t)$, demonstrating the viability of atom-based FM reception for higher baseband  bandwidths (here, $\sim 20~$Hz).

\begin{figure} [thb]
\vspace{1ex}
\centering
\includegraphics[width=0.4\textwidth]{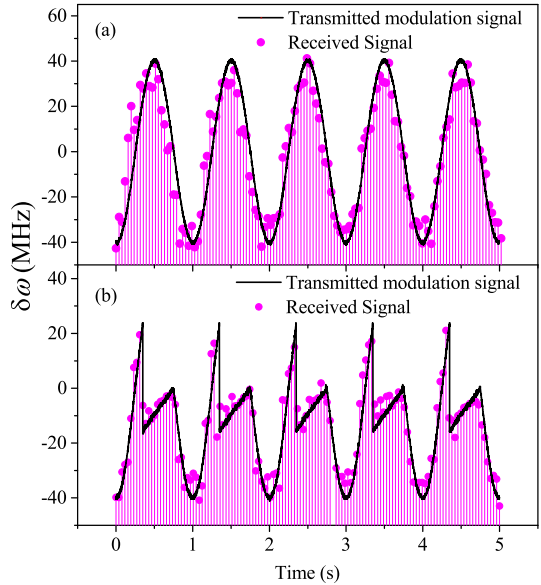}
\caption{Comparisons of transmitted and received FM signals, $\delta \omega_{FM}(t)$ (black lines) and $\delta \omega_{Rec}(t)$ (symbols with vertical drop lines) for (a) sinusoidal and (b) arbitrary FM modulation. The modulation frequency is 2$\pi \times$1~Hz, the MW Rabi frequency used in Eq.~\ref{eq:FM} is $\Omega=2\pi \times 60$~MHz, and the maximum absolute values of $\delta \omega$ are $\approx 2 \pi \times$40~MHz. }
\end{figure}

\section{Discussion and conclusion}

Figures 2, 3 and 5 demonstrate that Rydberg EIT-AT in an atomic vapor can be employed as a quantum sensor that enables real-time and direct readout of AM and FM baseband signals modulated onto electromagnetic carrier waves in the GHz range.
For detecting AM-modulated baseband signals, we employ the fact that on-resonant AT splittings depend linearly on the MW amplitude, as seen in Eq.~\ref{eq:AT}. For the AM test signals we use in Figs.~2(b), 3(a) and 3(b), whose complexity is increasing in that order,
we find average deviations between the transmitted and retrieved AM signal functions of 5.0\%, 1.6\% and 5.7\%, corresponding to fidelities of
95\% and 98.4\% and 94.3\%.

In FM modulation the Rydberg EIT-AT spectra are asymmetric, with the height difference between the two AT peaks providing a direct readout for the FM baseband signal. In our demonstration for a sinusoidal and an arbitrary baseband signal, the respective deviations of the readout from the baseband signal are about $3.0\%$ and $6.3\%$, corresponding to fidelities  of about $97\%$ and $93.7\%$.

In both the AM and FM case, the deviations are mostly attributed to the limited sampling rate of data-taking system, the laser frequency jitter, and the intrinsic EIT linewidth. The nonlinear dependence of the AT splitting on the microwave Rabi frequency~\cite{C. Holloway2017} may also affect the fidelity.

Our work paves the way towards the employment of atom-based quantum sensors and receivers in broadband atom-based microwave communications. The method could outperform traditional communications techniques that involve large receiver antennas.
The signal recovery from the atom-based spectroscopic data does not require an electronic demodulation process.
The range of carrier frequencies that can, in principle, be employed is quite vast. This is because strong electric-dipole transitions between conveniently accessible Rydberg levels span a frequency range from $\lesssim 1$~GHz up to $\sim 1$~THz. The signal bandwidth in the baseband used in this work is $\sim 10$~Hz. It is expected that this can be increased to several kHz using a faster data acquisition method for the spectroscopic data.

The work was supported by the National Key R$\&$D Program of China (Grant No. 2017YFA0304203), the National Natural Science
Foundation of China (Grant Nos. 61475090, 61675123, 61775124), the Changjiang Scholars and Innovative Research Team
in University of Ministry of Education of China (Grant No. IRT13076), the Key Program of the National Natural Science
Foundation of China (Grant No. 11434007) and Shanxi ¡°1331 Project¡± Key Subjects Construction. GR acknowledges support by the NSF (PHY-1506093) and BAIREN plan of Shanxi province.

\end{document}